\documentstyle[amsmath,amsfonts]{paper}
\def\be{\begin{equation}}
\def\ee{\end{equation}}
\def\ba{\begin{array}}
\def\ea{\end{array}}
\def\bea{\begin{eqnarray}}
\def\eea{\end{eqnarray}}
\def\e{{\rm e}}
\def\Exp{{\rm Exp}}

\def\d{{\rm d}}

\def\Det{{\rm Det}}

\bibliography{plain}
\title{Differential geometry on $SU(N)$: Left and right invariant vector fields and one-forms } \vspace{20mm}
\author{
  S. J. Akhtarshenas
\thanks{E-mail:akhtarshenas@phys.ui.ac.ir}
\\
{\small Department of Physics, University of Isfahan, Hezar Jerib
Ave., Isfahan, Iran }\\
{\small Quantum Optics Group, University of Isfahan, Hezar Jerib
Ave., Isfahan, Iran } }
\begin{document}
\maketitle
%\vspace{15mm}
%\newpage

\begin{abstract}
In this paper we provide an analytical procedure for explicit
calculation of the left and right invariant vector fields and
one-forms on $SU(N)$ manifold. The calculations are  based on the
coset parametrization of $SU(N)$ group. The results enable us to
calculate the invariant measure or Haar measure on the group. As
an illustrative example, we calculate invariant vector fields and
one-forms on $SU(2)$ group.


\end{abstract}
%\pagebreak
%\vspace{7cm}
\section{Introduction}
Because of the various applications of the group of unitary
transformation in physics, there is a great deal of attention in
investigation of the properties of the unitary group $SU(N)$. In
view of such considerable interest a lot of work has been devoted
to describe and parameterize $SU(N)$ manifold. A generalized Euler
angle parametrization for $SU(N)$ and $U(N)$ groups is given by
Tilma and Sudarshan \cite{tilma3,tilma4}. Di\c{t}\v{a} has
provided a parametrization of unitary matrices based on the
factorization of $N\times N$ unitary matrices \cite{dita1,dita2}.
Using this parametrization he has provided an explicit
parametrization for general $N$-dimensional Hermitian operators
that may be considered either as Hamiltonian or density matrices
\cite{dita3}. The subgroups and the coset spaces of the $SU(3)$
group are also listed in \cite{Khanna1997} along with a discussion
of the geometry of the group manifold which is relevant to the
understanding of the geometric phase.

The differential geometry on unitary groups is also an important
task in theoretical physics. To achieve this  it is important to
having a parametrization to construct differential geometry for
any unitary group. Byrd \cite{Byrd1998} has calculated the left
and right invariant vector fields and one-forms on $SU(3)$ group.
His calculation is based on the Euler angle parametrization for
the $SU(3)$ group. He has used the invariant one-forms on $SU(3)$
group, and has studied the geometric phase over the space of
$3$-level quantum systems. On the other hand, based on the Euler
angle parametrization of $SU(3)$ group and the result of
\cite{Byrd1998}, Panahi et al \cite{Panahi2001} have obtained a
two-dimensional Hamiltonian on $S^2$ via Fourier transformation
over the three coordinates of the $SU(3)$ Casimir operator defined
on $SU(3)/SU(2)$. Also by using the parametrization of
$SU(3)/SU(2)$ given in \cite{Khanna1997}, they have constructed
right invariant vector fields and the Casimir operator on
symmetric space $SU(3)/SU(2)$ and have obtained the
two-dimensional Hamiltonian of a charged particle on $S^2$ in the
presence of a an electric field \cite{Panahi2005}.

In this paper we present an analytical procedure for calculation
of the left and right invariant vector fields and one-forms on
$SU(N)$ group. This calculation is based on the coset
parametrization of $SU(N)$.  We also use the possibility of
factorizing each coset component in terms of a diagonal phase
matrix and an orthogonal matrix \cite{dita2,chaturvedi}. By using
this coset parametrization, we have recently given an explicit
expression for the Bures metric over the space of three-level
\cite{akhtar1} and $N$-level \cite{akhtar2} quantum systems. This
parametrization is convenient for many calculations, in the sense
that by using the coset parametrization the calculation can be
done in a unique manner for every $N$. Furthermore the coset
spaces appear in physics in several contexts, and provide an
elegant way of reducing a high-dimensional theory to a
lower-dimensional one. This paper, therefore, can be regarded as a
further development in the explicit calculation of the
differential geometric structure of $SU(N)$. The results of the
paper can have applications in studying the geometric phase over
the space of $N$-level quantum systems, and also in the context of
constructing Hamiltonian on some coset spaces of $SU(N)$ group.

The paper is organized as follows: In section 2, we review briefly
the Lie algebra of $SU(N)$ and introduce the generalized Gell-Mann
matrices as generators of the algebra.  The coset parametrization
for $SU(N)$ is introduced in section 3. In this section we also
obtain a formula for the volume of this group. Based on the coset
parametrization, we provide in section 4, a method to construct
the left and right invariant vector fields and one-forms on
$SU(N)$ manifold. In this section we also obtain the invariant
measure or Haar measure on the group. As an illustrative example
we obtain the differential geometry on $SU(2)$ in section 5. The
paper is concluded in section 6 with a brief conclusion.

\section{Preliminary: Lie algebra of $SU(N)$}
The group $SU(N)$ of $N\times N$ unitary matrices with unit
determinant is generated by the $N^2-1$ Hermitian, traceless
$N\times N$ matrices, that make the basis for the corresponding
Lie algebra $su(N)$. By choosing the normalization condition
$\textmd{Tr}(T_iT_j)=\frac{1}{2}\delta_{ij}$ for generators, we
can write the $(N-1)$ diagonal generators
$\{L_{\alpha_1}^{(1)}\}_{\alpha_1=1}^{N-1}$, i.e. Cartan
subalgebra, as (\cite{georgi}, page 114)
\begin{equation}\label{cartan}
(L_{\alpha_1}^{(1)})_{k,l}=\frac{1}{\sqrt{2\alpha_1(\alpha_1+1)}}\left(\sum_{j=1}^{\alpha_1}\delta_{k,j}\delta_{l,j}
-\alpha_1\delta_{k,\alpha_1+1}\delta_{l,\alpha_1+1}\right),
\end{equation}
and the  remaining $N(N-1)$ non-diagonal generators
$\{L_{\alpha_m}^{(m)}\}_{\alpha_m=1}^{2(m-1)}$ (for
$m=2,\cdots,N$) as follows
\begin{equation}\label{L}
(L_{\alpha_m}^{(m)})_{k,l}=\left\{\begin{array}{lll}\frac{1}{2}\left(\delta_{\alpha_m,
k}\delta_{m ,l}+\delta_{\alpha_m,l}\delta_{m, k}\right) & \rm{for}
& \alpha_m=1,\cdots,m-1
\\ \\
\frac{-i}{2}\left(\delta_{\alpha_m-m+1, k}\delta_{m,
l}-\delta_{\alpha_m-m+1, l}\delta_{m, k}\right) & \rm{for} &
\alpha_m=m,\cdots,2(m-1)
\end{array}\right. .
\end{equation}In the following sections, we set the range of indexes as
$\alpha_1=1,\cdots,N-1$ and $\alpha_m=1,\cdots,2(m-1)$ for
$m=2,\cdots, N$. The above $su(N)$ basis is the generalized
Gell-Mann matrices, and therefore, for the case of $N=2$ we get
the Pauli  matrices
$$
L_1^{(1)}=\left(\begin{array}{cc} 1 & 0 \\
0 & -1\end{array}\right),\qquad
L_1^{(2)}=\frac{1}{2}\left(\begin{array}{cc} 0 & 1 \\
1 & 0\end{array}\right),\qquad
L_2^{(2)}=\frac{1}{2}\left(\begin{array}{cc} 0 & -i \\
i& 0\end{array}\right).
$$
Also for $N=3$ we get the usual Gell-Mann matrices
$$
L_1^{(1)}=\frac{1}{2}\left(\begin{array}{ccc} 1 & 0 & 0\\
0 & -1 & 0 \\ 0 & 0 & 0\end{array}\right),\qquad
L_2^{(1)}=\frac{1}{2\sqrt{3}}\left(\begin{array}{ccc} 1 & 0 & 0\\
0 & 1 & 0 \\ 0 & 0 & -2\end{array}\right),
$$
$$
L_1^{(2)}=\frac{1}{2}\left(\begin{array}{ccc} 0 & 1 & 0\\
1 & 0 & 0 \\ 0 & 0 & 0\end{array}\right),\qquad
L_2^{(2)}=\frac{1}{2}\left(\begin{array}{ccc} 0 & -i & 0\\
i & 0 & 0 \\ 0 & 0 & 0\end{array}\right),
$$
$$
L_1^{(3)}=\frac{1}{2}\left(\begin{array}{ccc} 0 & 0 & 1\\
0 & 0 & 0 \\ 1 & 0 & 0\end{array}\right),\qquad
L_2^{(3)}=\frac{1}{2}\left(\begin{array}{ccc} 0 & 0 & 0\\
0 & 0 & 1 \\ 0 & 1 & 0\end{array}\right),
$$
$$
L_3^{(3)}=\frac{1}{2}\left(\begin{array}{ccc} 0 & 0 & -i\\
0 & 0 & 0 \\ i & 0 & 0\end{array}\right),\qquad
L_4^{(3)}=\frac{1}{2}\left(\begin{array}{ccc} 0 & 0 & 0\\
0 & 0 & -i \\ 0 & i & 0\end{array}\right).
$$

\section{Canonical coset parametrization of $SU(N)$ group}
In this section we provide a parametrization for $SU(N)$ group
that will be useful in calculating the differential geometry on
the group manifold. The parametrization is based on the coset
decomposition of unitary matrices.

\subsection{Coset factorization of $SU(N)$ group}
For every element $U\in SU(N)$, there is a unique decomposition of
$U$ into a product of $N$ group elements as \cite{gilmore}
\begin{equation}\label{Udecomposition}
U=\Omega^{(N;N)} \Omega^{(N-1;N)}\cdots
\Omega^{(2;N)}\Omega^{(1;N)}.
\end{equation}
In the above factorization we have
\begin{equation}
\Omega^{(1;N)}\in T^{N-1}
\end{equation}
where the $(N-1)$-dimensional torus $T^{N-1}$ is the product of
$N-1$ spheres $S^1=T^1$. A typical element for $\Omega^{(1;N)}$
can be represented as
\begin{equation}
\Omega^{(1;N)}= \Exp(i \eta_1 L_1^{(1)})\;\Exp(i \eta_{2}
L_{2}^{(1)})\cdots \Exp(i \eta_{N-1} L_{N-1}^{(1)}),
\end{equation}
where $\eta_{\alpha_1}$ for $\alpha_1=1,\cdots,N-1$ are real
parameters and $L_{\alpha_1}^{(1)}$ are Cartan generators defined
in equation (\ref{cartan}). The explicit form for $\Omega^{(1;N)}$
can be expressed as
\begin{equation}
(\Omega^{(1;N)})_{k,l}=\delta_{k,l}\Exp\left(-i\sqrt{\frac{k-1}{2k}}\eta_{k-1}
 +i\sum_{j=k}^{N-1}\frac{\eta_j}{\sqrt{2j(j+1)}}\right),
\end{equation}
with $\eta_0=0$. Also in decomposition (\ref{Udecomposition}) we
have the cosets
\begin{equation}
\Omega^{(m;N)}\in \frac{U(m)\otimes T^{N-m}}{U(m-1)\otimes
T^{N-m+1}}, \qquad m=2,\cdots,N.
\end{equation}
A typical coset representative $\Omega^{(m;N)}$ can be written as
\begin{equation}
\Omega^{(m;N)}= \left(
\begin{array}{c|c}
SU(m)/U(m-1) & O \\  \hline  O^T & I_{N-m}
\end{array}
\right),
\end{equation}
where $O$ represents the $m\times (N-m)$ zero matrix,  $O^T$ is
its transpose and $I_{N-m}$ denotes the unit matrix of order
$(N-m)$. The $2(m-1)$-dimensional coset space $SU(m)/U(m-1)$ has
the following $m\times m$ matrix representation (\cite{gilmore},
page 351)
\begin{equation}\label{cosetB}
SU(m)/U(m-1)= \left(
\begin{array}{c|c}
\cos{\sqrt{B^{(m)} [B^{(m)}]^\dag}} &
B^{(m)}\frac{\sin{\sqrt{[B^{(m)}]^{\dag}
B^{(m)}}}}{\sqrt{[B^{(m)}]^{\dag}B^{(m)}}} \\  \hline
-\frac{\sin{\sqrt{[B^{(m)}]^{\dag}
B^{(m)}}}}{\sqrt{[B^{(m)}]^{\dag}B^{(m)}}}[B^{(m)}]^\dag
&\cos{\sqrt{[B^{(m)}]^{\dag} B^{(m)}}} \\
\end{array}
\right),
\end{equation}
where $B^{(m)}$ represents an $(m-1)\times 1$ complex matrix and
$[B^{(m)}]^\dag$ is its adjoint.

\subsection{Factorization of coset  $\Omega^{(m;N)}$ }
Now by parameterizing the complex vector $B^{(m)}$ as
$(\gamma^{(m)}_1{\rm e}^{i\xi^{(m)}_1},\gamma^{(m)}_2{\rm
e}^{i\xi^{(m)}_2},\cdots,\gamma^{(m)}_{m-1}{\rm
e}^{i\xi^{(m)}_{m-1}})^T$ for $m=2,\cdots,N$, where
$\gamma^{(m)}_{i}$ and $\xi^{(m)}_{i}$ are real numbers, the
component $\Omega^{(m;N)}$ can be factorized as
\begin{equation}\label{OXRXd}
\Omega^{(m;N)}=X^{(m;N)}R^{(m;N)}{X^{(m;N)}}^\dag \qquad {\rm for}
\quad m=2,3,\cdots,N,
\end{equation}
where $X^{(m;N)}$ is a  diagonal  $N\times N$ phase matrix with
$X^{(m;N)}_{k,l}=\delta_{k,l}{\rm exp}\{i\xi^{(m)}_k\}$ and
$\xi^{(m)}_i=0$ for $i\ge m$, and $R^{(m;N)}$ is an $N\times N$
orthogonal matrix with the following nonzero elements
$$
R^{(m;N)}_{i,j}=\delta_{i,j} + {\hat \gamma}^{(m)}_i{\hat
\gamma}^{(m)}_j (\cos{\gamma^{(m)}}-1) \qquad {\rm for} \quad 1\le
i,j\le m-1
$$\vspace{-2mm}
$$ \hspace{-9mm}
R^{(m;N)}_{i,m}=-R^{(m;N)}_{m,i}={\hat \gamma}^{(m)}_i
\sin{\gamma^{(m)}}  \qquad{\rm for} \quad 1\le i\le m-1
$$\vspace{-2mm}
$$\hspace{-71mm}
R^{(m;N)}_{m,m}=\cos{\gamma^{(m)}}
$$\vspace{-2mm}
\begin{equation}\label{RmN}\hspace{-44mm}
R^{(m;N)}_{i,i}=1   \qquad{\rm for} \quad m+1\le i\le N,
\end{equation}
where we have defined ${\hat
\gamma}^{(m)}_i=\gamma^{(m)}_i/\gamma^{(m)}$ and
$\gamma^{(m)}=\sqrt{[B^{(m)}]^{\dag}
B^{(m)}}=\sqrt{\sum_{i=1}^{m-1}(\gamma^{(m)}_i)^2}$. As we will
see later the important ingredient of our approach in computing
the vector fields and one-forms is the possibility of writing the
factorization (\ref{OXRXd}).

\section{Differential geometry on $SU(N)$ }
In this section we will find differential operators corresponding
to the generators  of $SU(N)$ group. To this aim, we first define
 $\chi^{(m)}_{\alpha_m}$
$(\alpha_m=1,2,\cdots,2(m-1))$ as the $2(m-1)$ real parameters of
the coset component $\Omega^{(m;N)}$ $(m=2,\cdots,N)$ such that
\begin{equation}
\chi^{(m)}_{\alpha_m}=\left\{\begin{array}{ll}
\gamma^{(m)}_{\alpha_m} & {\rm for} \quad \alpha_m=1,\cdots,m-1
\\ \\ \xi^{(m)}_{\alpha_{m}-m+1} & {\rm for} \quad
\alpha_m=m,\cdots,2(m-1).
\end{array}\right.
\end{equation}

\subsection{Left invariant vector fields and one-forms}
Now in order to calculate left invariant vector fields, we first
take derivatives of the group element
$U(\eta_{\alpha_1},\chi^{(m)}_{\alpha_m})$
 with respect to each set of parameters $\eta_{\alpha_1}$ and
$ \chi^{(m)}_{\alpha_m}$, and write the result of differentiation
as
 \begin{equation}\label{}
    \frac{\partial}{\partial\eta_{\alpha_1}}U=UA_{\alpha_1}^{(1)},\qquad
    \frac{\partial}{\partial\chi_{\alpha_m}^{(m)}}U=UA_{\alpha_m}^{(m)},
\end{equation}
where the matrices $A_{\alpha_1}^{(1)}$ and $A_{\alpha_m}^{(m)}$
are define by
\begin{eqnarray}
A_{\alpha_1}^{(1)}&=&{\Omega^{(1,N)}}^\dag\frac{\partial\Omega^{(1,N)}}{\partial\eta_{\alpha_1}}=iL_{\alpha_1}^{(1)}
\\
A_{\alpha_m}^{(m)}&=&{W^{(m,N)}}^\dag
    \left(\frac{\partial\Omega^{(m,N)}}{\partial\chi^{(m)}_{\alpha_m}}
    {\Omega^{(m,N)}}^\dag\right)W^{(m,N)}
\end{eqnarray}
with
\begin{equation}
W^{(m;N)}=\Omega^{(m;N)}\Omega^{(m-1;N)}\cdots\Omega^{(2;N)}\Omega^{(1;N)}
\end{equation}
and the anti Hermitian matrix
$\left(\frac{\partial\Omega^{(m,N)}}{\partial\gamma^{(m)}_{\alpha_m}}
    {\Omega^{(m,N)}}^\dag\right)$ has the following nonzero matrix elements
\begin{equation}\label{antiH}
    \begin{array}{ll}
    \left(\frac{\partial\Omega^{(m,N)}}{\partial\gamma^{(m)}_{\alpha_m}}
    {\Omega^{(m,N)}}^\dag\right)_{r<m,s<m}
       & = \e^{i(\xi^{(m)}_r-\xi^{(m)}_s)}\left({\hat
       \gamma}^{(m)}_r\delta_{s,\alpha_m}-{\hat
       \gamma}^{(m)}_s\delta_{r,\alpha_m}
       \right)\left(\frac{\cos{\gamma}-1}{\gamma}\right)
       \\
      \left(\frac{\partial\Omega^{(m,N)}}{\partial\gamma^{(m)}_{\alpha_m}}
    {\Omega^{(m,N)}}^\dag\right)_{r<m,m} & =
    \e^{i\xi^{(m)}_r}\left({\hat
       \gamma}^{(m)}_r{\hat
       \gamma}^{(m)}_{\alpha_m}+\left(\delta_{r,\alpha_m}-{\hat
       \gamma}^{(m)}_r{\hat
       \gamma}^{(m)}_{\alpha_m}\right)\frac{\sin{\gamma}}{\gamma}
       \right) \\
      \left(\frac{\partial\Omega^{(m,N)}}{\partial\gamma^{(m)}_{\alpha_m}}
    {\Omega^{(m,N)}}^\dag\right)_{m,r<m} & =
    -\left(\frac{\partial\Omega^{(m,N)}}{\partial\gamma^{(m)}_{\alpha_m}}
    {\Omega^{(m,N)}}^\dag\right)_{r<m,m}^\ast \\
      \left(\frac{\partial\Omega^{(m,N)}}{\partial\gamma^{(m)}_{\alpha_m}}
    {\Omega^{(m,N)}}^\dag\right)_{m,m} & =0, \\
    \end{array}
\end{equation}
and also the matrix
$\left(\frac{\partial\Omega^{(m;N)}}{\partial\xi^{(m)}_{\alpha_m}}
    {\Omega^{(m;N)}}^\dag\right)$ defined by
 \begin{equation}\label{}
    \left(\frac{\partial\Omega^{(m;N)}}{\partial\xi^{(m)}_{\alpha_m}}
    {\Omega^{(m;N)}}^\dag\right)_{r,s}=i\delta_{r,\alpha_m}\delta_{s,\alpha_m}-
    iR^{(m;N)}_{r,\alpha_m}R^{(m;N)}_{s,\alpha_m}\e^{i(\xi^{(m)}_r-\xi^{(m)}_s)}.
\end{equation}
The matrices $A_{\alpha_1}^{(1)}$ and $A_{\alpha_m}^{(m)}$  can be
expanded as a linear combination of the Lie algebra basis
(\ref{cartan}) and (\ref{L}) as follows
\begin{eqnarray}\label{}
        A_{\alpha_1}^{(1)}&=&\sum_{{\beta_1}=1}^{N-1}a_{{\alpha_1},{\beta_1}}^{(1,1)}
        L_{\beta_1}^{(1)}
    +\sum_{{m^\prime}=2}^{N}\sum_{\beta_{m^\prime}=1}^{{2(m^\prime}-1)}
    a_{{\alpha_1},\beta_{m^\prime}}^{(1,{m^\prime})}L_{\beta_{m^\prime}}^{({m^\prime})},
\\
        A_{\alpha_m}^{(m)}&=&\sum_{{\beta_1}=1}^{N-1}a_{\alpha_m,{\beta_1}}^{(m,1)}
        L_{\beta_1}^{(1)}
    +\sum_{m^\prime=2}^{N}\sum_{\beta_{m^\prime=1}}^{2(m^\prime-1)}
    a_{\alpha_m,\beta_{m^\prime}}^{(m,m^\prime)}L_{\beta_{m^\prime}}^{(m^\prime)}.
\end{eqnarray}
The coefficients of the above expansion can be obtained from the
orthogonality of the generators of the Lie algebra. We obtain the
following results
\begin{eqnarray}
a_{\alpha_1,\beta_1}^{(1,1)}&=&i\delta_{\alpha_1,\beta_1}
\\
a_{\alpha_1,\beta_m}^{(1,m)}&=& 0
\\
    a_{\alpha_m,\beta_1}^{(m,1)}&=&
    \frac{2}{\sqrt{2\beta_1(\beta_1+1)}}
    \left(\sum_{j=1}^{\beta_1}\left(A_{\alpha_m}^{(m)}\right)_{jj}-
    \beta_1\left(A_{\alpha_m}^{(m)}\right)_{\beta_1+1,\beta_1+1}
    \right)
\\
    a_{\alpha_{m},\beta_{m'}}^{(m,m')}&=&
    \left\{\begin{array}{lll}
    \left(\left(A_{\alpha_m}^{(m)}\right)_{m',\beta_{m'}}+
    \left(A_{\alpha_m}^{(m)}\right)_{\beta_{m'},m'}
    \right) & \rm{for} & \beta_{m'}=1,\cdots,m-1
\\
    -i\left(\left(A_{\alpha_m}^{(m)}\right)_{m',\beta_{m'}}-
    \left(A_{\alpha_m}^{(m)}\right)_{\beta_{m'},m'}
    \right) & \rm{for} & \beta_{m'}=m',\cdots,2(m-1)
    \end{array}\right.
\end{eqnarray}
Therefore by choosing the following order for coordinates
\begin{equation}\label{order}
\{\eta_{1},\cdots,\eta_{N-1}\};
\{\gamma_{1}^{(2)},\xi_{1}^{(2)}\}; \cdots;
\{\gamma_{1}^{(N)},\cdots,\gamma_{N-1}^{(N)},\xi_{1}^{(N)},\cdots,\xi_{N-1}^{(N)}\}.
\end{equation}
we can define the matrix $A$ by
\begin{equation}\label{}
    A=\left(%
\begin{array}{c|c|c|c|c}
  a_{\alpha_1,\beta_1}^{(1,1)} & a_{\alpha_1,\beta_2}^{(1,2)}
  & a_{\alpha_1,\beta_3}^{(1,3)} & \cdots &  a_{\alpha_1,\beta_N}^{(1,N)}\\
  \hline
  a_{\alpha_2,\beta_1}^{(2,1)} & a_{\alpha_2,\beta_2}^{(2,2)}
  & a_{\alpha_2,\beta_3}^{(2,3)} & \cdots &  a_{\alpha_2,\beta_N}^{(2,N)}\\
  \hline
  a_{\alpha_3,\beta_1}^{(3,1)} & a_{\alpha_3,\beta_2}^{(3,2)}
  & a_{\alpha_3,\beta_3}^{(3,3)} & \cdots &  a_{\alpha_3,\beta_N}^{(3,N)}\\
   \hline
  \vdots & \vdots
  & \vdots & \ddots &  \vdots\\
  \hline
  a_{\alpha_N,\beta_1}^{(N,1)} & a_{\alpha_N,\beta_2}^{(N,2)}
  & a_{\alpha_N,\beta_3}^{(N,3)} & \cdots &  a_{\alpha_N,\beta_N}^{(N,N)}\\
\end{array}%
\right).
\end{equation}
With the help of this matrix the  differential operators are
related to the Lie algebra generators as follows
\begin{equation}\label{}
    \left(%
\begin{array}{c}
  \partial/\partial\eta_{\alpha_1}\\
   \\
 \partial/\partial\chi_{\alpha_m}^{(m)}
\end{array}%
\right)=A
\left(%
\begin{array}{c}
  L_{\beta_1}^{(1)} \\ \\
  L_{\beta_{m^\prime}}^{(m^\prime)}
\end{array}%
\right).
\end{equation}
Now by tacking inverse of the matrix $A$, the left invariant
vector fields on $SU(N)$ group can be obtained  by the following
equation
\begin{equation}\label{}
\left(\begin{array}{c}
  \Lambda_{\beta_1}^{(1)} \\ \\
  \Lambda_{\beta_{m^\prime}}^{(m^\prime)}
\end{array}
\right) =A^{-1}
\left(%
\begin{array}{c}
  \partial/\partial\eta_{\alpha_1}\\
   \\
 \partial/\partial\chi_{\alpha_m}^{(m)}
\end{array}\right).
\end{equation}

Now in order to calculate the left invariant one-forms, we first
expand them in terms of the basis $\d \eta_{\beta_1}$ and $\d
\chi^{(m)}_{\beta_{m}}$

\begin{eqnarray}\label{}
   \omega_{\alpha_1}^{(1)}&=&\sum_{\beta_1=1}^{N-1}c_{\alpha_1,\beta_1}^{(1,1)}\d
\eta_{\beta_1}
+\sum_{m^\prime=2}^{N}\sum_{\beta_{m^\prime}=1}^{2(m^\prime-1)}c_{\alpha_1,\beta_{m^\prime}}^{(1,m^\prime)}\d
\chi^{(m^\prime)}_{\beta_{m^\prime}},
\\
        \omega_{\alpha_m}^{(m)}&=&\sum_{{\beta_1}=1}^{N-1}c_{\alpha_m,{\beta_1}}^{(m,1)}
        \d \eta_{\beta_1}
    +\sum_{m^\prime=2}^{N}\sum_{\beta_{m^\prime=1}}^{2(m^\prime-1)}
    c_{\alpha_m,\beta_{m^\prime}}^{(m,m^\prime)}\d
    \chi_{\beta_{m^\prime}}^{(m^\prime)}.
\end{eqnarray}
By using the fact that left invariant one-forms are dual to the
left invariant vector fields, i.e.
$$
\begin{array}{ll}
(\omega_{\alpha_1}^{(1)},\Lambda_{{\alpha^\prime}_1}^{(1)})=\delta_{\alpha_1,{\alpha^\prime}_1}
& (\omega_{\alpha_1}^{(1)},\Lambda_{{\alpha}_m}^{(m)})=0
\\
(\omega_{\alpha_{\alpha_m}}^{(m)},\Lambda_{{\alpha^\prime}_1}^{(1)})=0
&
(\omega_{\alpha_m}^{(m)},\Lambda_{{\alpha'}_{m'}}^{(m')})=\delta_{\alpha_m,{\alpha^\prime}_{m'}}\delta_{m,
m'}, \end{array}
$$
we find that $ C^{T}(A^{-1})=\mathbb{I}$ and therefore, we have
$C=A^{T}$. Consequently, the left invariant one-forms are obtained
by tacking the transpose of the matrix $A$, i.e.
\begin{equation}\label{}
\left(\begin{array}{c}
  \omega_{\beta_1}^{(1)} \\ \\
  \omega_{\beta_{m^\prime}}^{(m^\prime)}
\end{array}
\right) =A^{T}
\left(%
\begin{array}{c}
  \d\eta_{\alpha_1}\\
   \\
 \d\chi_{\alpha_m}^{(m)}
\end{array}\right).
\end{equation}

\subsection{Right invariant vector fields and one-forms}

Now in order to calculate right invariant vector fields, we first
take derivatives of the group element
$U(\eta_{\alpha_1},\chi^{(m)}_{\alpha_m})$
 with respect to each set of parameters $\eta_{\alpha_1}$ and
$ \chi^{(m)}_{\alpha_m}$, and write the result of differentiation
as
 \begin{equation}\label{}
    \frac{\partial}{\partial\eta_{\alpha_1}}U={\widetilde A}_{\alpha_1}^{(1)}U,\qquad
    \frac{\partial}{\partial\chi_{\alpha_m}^{(m)}}U={\tilde A}_{\alpha_m}^{(m)}U,
\end{equation}
where the matrices ${\widetilde A}_{\alpha_1}^{(1)}$ and
${\widetilde A}_{\alpha_m}^{(m)}$ are define by
\begin{eqnarray}
{\widetilde A}_{\alpha_1}^{(1)}&=&{{\widetilde W}^{(1,N)}}
\left(\frac{\partial\Omega^{(1,N)}}{\partial\eta_{\alpha_1}}{\Omega^{(1,N)}}^\dag\right){\widetilde
W}^{(1,N)^\dag}
\\
{\widetilde A}_{\alpha_m}^{(m)}&=&{{\widetilde W}^{(m,N)}}
\left(\frac{\partial\Omega^{(m,N)}}{\partial\chi^{(m)}_{\alpha_m}}
    {\Omega^{(m,N)}}^\dag\right){\widetilde W}^{(m,N)^{\dag}},
\end{eqnarray}
with
\begin{equation}
{\widetilde
W}^{(k;N)}=\Omega^{(N;N)}\Omega^{(N-1;N)}\cdots\Omega^{(k+1;N)}=U{W^{(k;N)}}^\dag,
\end{equation}
and
$$
\left(\frac{\partial\Omega^{(1,N)}}{\partial\eta_{\alpha_1}}{\Omega^{(1,N)}}^\dag\right)=
\left({\Omega^{(1,N)}}^\dag\frac{\partial\Omega^{(1,N)}}{\partial\eta_{\alpha_1}}\right)=iL_{\alpha_1}^{(1)},
$$
and the anti Hermitian matrix
$\left(\frac{\partial\Omega^{(m,N)}}{\partial\gamma^{(m)}_{\alpha_m}}
    {\Omega^{(m,N)}}^\dag\right)$ has defined in equation (\ref{antiH}).

The matrices ${\widetilde A}_{\alpha_1}^{(1)}$ and ${\widetilde
A}_{\alpha_m}^{(m)}$ can be expanded as a linear combination of
the Lie algebra
\begin{eqnarray}\label{}
        {\widetilde A}_{\alpha_1}^{(1)}&=&\sum_{{\beta_1}=1}^{N-1}{\tilde a}_{{\alpha_1},{\beta_1}}^{(1,1)}
        L_{\beta_1}^{(1)}
    +\sum_{{m^\prime}=2}^{N}\sum_{\beta_{m^\prime}=1}^{{2(m^\prime}-1)}
    {\tilde
    a}_{{\alpha_1},\beta_{m^\prime}}^{(1,{m^\prime})}L_{\beta_{m^\prime}}^{({m^\prime})},
\\
        {\widetilde A}_{\alpha_m}^{(m)}&=&\sum_{{\beta_1}=1}^{N-1}{\tilde a}_{\alpha_m,{\beta_1}}^{(m,1)}
        L_{\beta_1}^{(1)}
    +\sum_{m^\prime=2}^{N}\sum_{\beta_{m^\prime=1}}^{2(m^\prime-1)}
    {\tilde
    a}_{\alpha_m,\beta_{m^\prime}}^{(m,m^\prime)}L_{\beta_{m^\prime}}^{(m^\prime)}.
\end{eqnarray}
The coefficients of the above expansion can be obtained from the
orthogonality of the generators of the Lie algebra. We obtain the
following results
\begin{eqnarray}
{\tilde
a}_{\alpha_1,\beta_1}^{(1,1)}&=&\frac{2}{\sqrt{2\beta_1(\beta_1+1)}}
    \left(\sum_{j=1}^{\beta_1}\left({\widetilde A}_{\alpha_1}^{(1)}\right)_{jj}-
    \beta_1\left({\widetilde A}_{\alpha_1}^{(1)}\right)_{\beta_1+1,\beta_1+1}
    \right)
\\
{\tilde a}_{\alpha_1,\beta_m}^{(1,m)}&=& \left\{\begin{array}{lll}
    \left(\left({\widetilde A}_{\alpha_1}^{(1)}\right)_{m,\beta_{m}}+
    \left({\widetilde A}_{\alpha_1}^{(1)}\right)_{\beta_{m},m}
    \right) & \rm{for} & \beta_{m}=1,\cdots,m-1
\\
    -i\left(\left({\widetilde A}_{\alpha_1}^{(1)}\right)_{m,\beta_{m}}-
    \left({\widetilde A}_{\alpha_1}^{(1)}\right)_{\beta_{m},m}
    \right) & \rm{for} & \beta_{m}=m,\cdots,2(m-1) \end{array}\right.
\\
    {\tilde a}_{\alpha_m,\beta_1}^{(m,1)}&=&
    \frac{2}{\sqrt{2\beta_1(\beta_1+1)}}
    \left(\sum_{j=1}^{\beta_1}\left({\widetilde A}_{\alpha_m}^{(m)}\right)_{jj}-
    \beta_1\left({\widetilde A}_{\alpha_m}^{(m)}\right)_{\beta_1+1,\beta_1+1}
    \right)
\\
    {\tilde a}_{\alpha_{m},\beta_{m'}}^{(m,m')}&=&
    \left\{\begin{array}{lll}
    \left(\left({\widetilde A}_{\alpha_m}^{(m)}\right)_{m',\beta_{m'}}+
    \left({\widetilde A}_{\alpha_m}^{(m)}\right)_{\beta_{m'},m'}
    \right) & \rm{for} & \beta_{m'}=1,\cdots,m-1
\\
    -i\left(\left({\widetilde A}_{\alpha_m}^{(m)}\right)_{m',\beta_{m'}}-
    \left({\widetilde A}_{\alpha_m}^{(m)}\right)_{\beta_{m'},m'}
    \right) & \rm{for} & \beta_{m'}=m',\cdots,2(m-1) \end{array}\right.
\end{eqnarray}
By choosing the coordinates order given in (\ref{order}), we get
the following representation for matrix ${\widetilde A}$
\begin{equation}\label{}
   {\widetilde A}=\left(%
\begin{array}{c|c|c|c|c}
  {\tilde a}_{\alpha_1,\beta_1}^{(1,1)} & {\tilde a}_{\alpha_1,\beta_2}^{(1,2)}
  & {\tilde a}_{\alpha_1,\beta_3}^{(1,3)} & \cdots &  {\tilde a}_{\alpha_1,\beta_N}^{(1,N)}\\
  \hline
  {\tilde a}_{\alpha_2,\beta_1}^{(2,1)} & {\tilde a}_{\alpha_2,\beta_2}^{(2,2)}
  & {\tilde a}_{\alpha_2,\beta_3}^{(2,3)} & \cdots &  {\tilde a}_{\alpha_2,\beta_N}^{(2,N)}\\
  \hline
  {\tilde a}_{\alpha_3,\beta_1}^{(3,1)} & {\tilde a}_{\alpha_3,\beta_2}^{(3,2)}
  & {\tilde a}_{\alpha_3,\beta_3}^{(3,3)} & \cdots &  {\tilde a}_{\alpha_3,\beta_N}^{(3,N)}\\
   \hline
  \vdots & \vdots
  & \vdots & \ddots &  \vdots\\
  \hline
  {\tilde a}_{\alpha_N,\beta_1}^{(N,1)} & {\tilde a}_{\alpha_N,\beta_2}^{(N,2)}
  & {\tilde a}_{\alpha_N,\beta_3}^{(N,3)} & \cdots &  {\tilde a}_{\alpha_N,\beta_N}^{(N,N)}\\
\end{array}%
\right),
\end{equation}
where can be used to write the differential operators in terms of
the Lie algebra generators as
\begin{equation}\label{}
    \left(%
\begin{array}{c}
  \partial/\partial\eta_{\alpha_1}\\
   \\
 \partial/\partial\chi_{\alpha_m}^{(m)}
\end{array}%
\right)={\widetilde A}
\left(%
\begin{array}{c}
  L_{\beta_1}^{(1)} \\ \\
  L_{\beta_{m^\prime}}^{(m^\prime)}
\end{array}%
\right).
\end{equation}
Consequently, the right invariant vector fields on the $SU(N)$
group can be obtained by the following
\begin{equation}\label{}
\left(\begin{array}{c}
  {\widetilde \Lambda}_{\beta_1}^{(1)} \\ \\
  {\widetilde\Lambda}_{\beta_{m^\prime}}^{(m^\prime)}
\end{array}
\right) ={\widetilde A}^{-1}
\left(%
\begin{array}{c}
  \partial/\partial\eta_{\alpha_1}\\
   \\
 \partial/\partial\chi_{\alpha_m}^{(m)}
\end{array}\right).
\end{equation}

Finally, similar to the case of left invariant one-forms by
defining ${\widetilde A}^{T}$ as the transpose of ${\widetilde
A}$, the right invariant one-forms can be written as
\begin{equation}\label{}
\left(\begin{array}{c}
  {\widetilde \omega}_{\beta_1}^{(1)} \\ \\
  {\widetilde \omega}_{\beta_{m^\prime}}^{(m^\prime)}
\end{array}
\right) ={\widetilde A}^{T}
\left(%
\begin{array}{c}
  \d\eta_{\alpha_1}\\
   \\
 \d\chi_{\alpha_m}^{(m)}
\end{array}\right).
\end{equation}

\subsection{Invariant integration measure } Invariant measure, or Haar measure,
on the $SU(N)$ manifold can be obtained by tacking the wedge
product of the left or right invariant one-forms. The result is,
of course, equivalent by tacking the determinant of the matrix $A$
(or matrix ${\widetilde A}$ because of the fact that left and
right invariant measures on $SU(N)$ are equal), where we get
$$
\d\mu[SU(N)]=\Det(A)\prod_{\alpha_1=1}^{N-1}\d\eta_{\alpha_1}
\prod_{m=2}^{N}\prod_{\alpha_m=1}^{m-1}\d\gamma_{\alpha_m}\d\xi_{\alpha_m}.
$$

\section{An example: The $SU(2)$ group}
In this section we consider for more illustration the case of
$N=2$ explicitly. In this particular case we have three parameters
$\eta_1$, $\gamma_1^{(2)}$ and $\xi_1^{(2)}$, where for the sake
of simplicity we set them as $\eta$, $\gamma$ and $\xi$
respectively. In this  case we see that
$$
\Omega^{(2;2)}=\left(\begin{array}{cc} \cos{\gamma} &
\e^{i\xi}\sin{\gamma} \\
-\e^{-i\xi}\sin{\gamma}  & \cos{\gamma}
\end{array}\right), \qquad
\Omega^{(1;2)}=\left(\begin{array}{cc}  \e^{i\eta/2} &  0\\
0 & \e^{-i\eta/2}
\end{array}\right),
$$
and therefore
$$
W^{(2,2)}=\left(\begin{array}{cc} \cos{\gamma}\e^{i\eta/2} &
\e^{i(\xi-\eta/2)}\sin{\gamma} \\
-\e^{-i(\xi-\eta/2)}\sin{\gamma}  & \cos{\gamma}\e^{-i\eta/2}
\end{array}\right),
$$
We  also find
$$
A_1^{(2)}=\left(\begin{array}{cc} 0 &
\e^{i(\xi-\eta)} \\
-\e^{-i(\xi-\eta)}  & 0
\end{array}\right),
\quad A_2^{(2)}=\left(\begin{array}{cc} -i\sin^2{\gamma} &
i\sin{\gamma}\cos{\gamma}\e^{i(\xi-\eta)} \\
i\sin{\gamma}\cos{\gamma}\e^{-i(\xi-\eta)} & i\sin^2{\gamma}
\end{array}\right).
$$
where can be used to write the matrix $A$ as
$$
A=\left(\begin{array}{c|cc} i & 0 & 0 \\  \hline 0  &
2i\sin{(\xi-\eta)} & 2i\cos{(\xi-\eta)}
\\
-i(1-\cos{2\gamma}) & i\sin{2\gamma}\cos{(\xi-\eta)} &
-i\sin{2\gamma}\sin{(\xi-\eta)}
\end{array}\right),
$$
with the following inverse
$$
A^{-1}=\left(\begin{array}{c|cc} -i & 0 & 0 \\  \hline
-i\cos{(\xi-\eta)}\tan{\gamma}  & -\frac{i}{2}\sin{(\xi-\eta)} &
-i\cos{(\xi-\eta)}\csc{2\gamma}
\\
i\sin{(\xi-\eta)}\tan{\gamma} & -\frac{i}{2}\cos{(\xi-\eta)} &
i\sin{(\xi-\eta)}\csc{2\gamma}
\end{array}\right).
$$
These two matrices can be used to obtain the left invariant vector
fields as
\begin{eqnarray}\label{}\nonumber
\Lambda_{1}^{(1)}&=&-i\frac{\partial}{\partial\eta}, \\
\nonumber\Lambda_{1}^{(2)}&=&-i\cos{(\xi-\eta)}\tan{\gamma}
\frac{\partial}{\partial\eta}  -\frac{i}{2}\sin{(\xi-\eta)}
\frac{\partial}{\partial\gamma}
-i\cos{(\xi-\eta)}\csc{2\gamma}\frac{\partial}{\partial\xi},
\\ \nonumber
\Lambda_{2}^{(2)}&=&i\sin{(\xi-\eta)}\tan{\gamma}
\frac{\partial}{\partial\eta}  -\frac{i}{2}\cos{(\xi-\eta)}
\frac{\partial}{\partial\gamma}
+i\sin{(\xi-\eta)}\csc{2\gamma}\frac{\partial}{\partial\xi},
\end{eqnarray}
and the left invariant one-forms  as follows
\begin{eqnarray}\label{}\nonumber
\omega_{1}^{(1)}&=&  i\d\eta -i(1-\cos{2\gamma})\d\xi, \\
\nonumber \omega_{1}^{(2)}&=& 2i\sin{(\xi-\eta)}\d\gamma
+i\sin{2\gamma}\cos{(\xi-\eta)} \d\xi,\\
\nonumber \omega_{2}^{(2)}&=& 2i\cos{(\xi-\eta)} \d\gamma
-i\sin{2\gamma}\sin{(\xi-\eta)}\d\xi.
\end{eqnarray}

For calculation of the right invariant vector fields we note that
in this particular case we have ${\widetilde
W}^{(1;2)}=\Omega^{(2;2)}$ and ${\widetilde W}^{(2;2)}=I$, and
therefore we find
$$
{\widetilde A}_1^{(1)}=\left(\begin{array}{cc}
\frac{i}{2}\cos{2\gamma} &
\frac{-i}{2}\sin{2\gamma}\e^{i\xi} \\
\frac{-i}{2}\sin{2\gamma}\e^{-i\xi}  & \frac{-i}{2}\cos{2\gamma}
\end{array}\right),
$$
$$
{\widetilde A}_1^{(2)}=\left(\begin{array}{cc} 0 &
\e^{i\xi} \\
-\e^{-i\xi}  & 0
\end{array}\right),
\quad {\widetilde A}_2^{(2)}=\left(\begin{array}{cc}
i\sin^2{\gamma} &
i\sin{\gamma}\cos{\gamma}\e^{i\xi} \\
-i\sin{\gamma}\cos{\gamma}\e^{-i\xi} & -i\sin^2{\gamma}
\end{array}\right).
$$
We then obtain the matrix ${\widetilde A}$  and its inverse
${\widetilde A}^{-1}$ as
$$
{\widetilde A}=\left(\begin{array}{c|cc} i\cos{2\gamma} &
-i\sin{2\gamma}\cos{\xi} &
i\sin{2\gamma}\sin{\xi} \\
\hline 0 & 2i\sin{\xi} & 2i\cos{\xi}
\\
i(1-\cos{2\gamma}) & i\sin{2\gamma}\cos{\xi} &
-i\sin{2\gamma}\sin{\xi}
\end{array}\right),
$$
and
$$
{\widetilde A}^{-1}=\left(\begin{array}{c|cc} -i & 0 & -i \\
\hline i\cos{\xi}\tan{\gamma}  & -\frac{i}{2}\sin{\xi_1^{(2)}} &
-i\cos{\xi_1^{(2)}}\cot{2\gamma}
\\
-i\sin{\xi}\tan{\gamma} & -\frac{i}{2}\cos{\xi} &
i\sin{\xi}\cot{2\gamma}
\end{array}\right),
$$
where can be used to obtain the right invariant vector fields
\begin{eqnarray}\label{}\nonumber
{\widetilde
\Lambda}_{1}^{(1)}&=&-i\frac{\partial}{\partial\eta}-i\frac{\partial}{\partial\xi},
\\ \nonumber
{\widetilde \Lambda}_{1}^{(2)}&=&i\cos{\xi}\tan{\gamma}
\frac{\partial}{\partial\eta} -\frac{i}{2}\sin{\xi}
\frac{\partial}{\partial\gamma_1^{(2)}}
-i\cos{\xi}\cot{2\gamma}\frac{\partial}{\partial\xi},
\\ \nonumber
{\widetilde \Lambda}_{2}^{(2)}&=&-i\sin{\xi}\tan{\gamma}
\frac{\partial}{\partial\eta} -\frac{i}{2}\cos{\xi}
\frac{\partial}{\partial\gamma_1^{(2)}}
+i\sin{\xi}\cot{2\gamma}\frac{\partial}{\partial\xi},
\end{eqnarray}
and the right invariant one-forms
\begin{eqnarray}\label{}\nonumber
{\widetilde \omega}_{1}^{(1)}&=&  i\cos{2\gamma}\d\eta +i(1-\cos{2\gamma})\d\xi, \\
\nonumber {\widetilde \omega}_{1}^{(2)}&=&
-i\sin{2\gamma}\cos{\xi}\d\eta +2i\sin{\xi}\d\gamma
+i\sin{2\gamma}\cos{\xi} \d\xi, \\
\nonumber {\widetilde \omega}_{2}^{(2)}&=& i\sin{2\gamma}\sin{\xi}
\d\eta +2i\cos{\xi}\d\gamma -i\sin{2\gamma}\sin{\xi}\d\xi.
\end{eqnarray}
Finally, the invariant measure of this group is also obtained as
$$
\d\mu[SU(2)]=2\sin{2\gamma}\d\eta\d\gamma\d\xi.
$$

\section{Conclusion}
In this paper we present a method for explicit calculation of the
left and right invariant vector fields and one-forms on $SU(N)$
manifold. The calculations are based on the coset parametrization
of the unitary group $SU(N)$.  It is shown that in the canonical
coset parametrization, the invariant measure on the group manifold
is decomposed as the product of the invariant measure on the
constitute cosets. The advantage of this approach is the
possibility of calculating  explicitly the differential geometry
on $SU(N)$ for every $N$, in the sense that by knowing $N$, it is
enough  to construct the matrix $A$ and then taking the transpose
and inverse of this matrix, which are not difficult task to
handel. As an illustrative example we calculate explicitly, the
differential structure on $SU(2)$ group.


\begin{thebibliography}{99}





\bibitem{tilma3}{ T. Tilma and E. C. G. Sudarshan,}
{\em J. Phys. A: Math. Gen. {\bf 35}, 10467 (2002).}
\bibitem{tilma4}{ T. Tilma and E. C. G. Sudarshan,}
{\em J. Geom. Phys. {\bf 52}, 263 (2004).}
\bibitem{dita1}{ P. Di\c{t}\v{a},}
{\em J. Phys. A: Math. Gen. {\bf 15}, 3465 (1982).}
\bibitem{dita2}{ P. Di\c{t}\v{a},}
{\em J. Phys. A: Math. Gen. {\bf 36}, 2781 (2003).}
\bibitem{dita3}{ P. Di\c{t}\v{a},}
{\em J. Phys. A: Math. Gen. {\bf 38}, 2657 (2005).}
\bibitem{Khanna1997}{ G. Khanna, S. Mukhopadhyay, R. Simon and N. Mukunda,}
{\em Ann. Phys. (N.Y.) {\bf 253}, 55 (1997).}
\bibitem{Byrd1998}{ M. S. Byrd,}
{\em J. Math. Phys. {\bf 39}, 6125 (1998), {\it ibid} {\bf 41},
1026 (2000) Erratum.}
\bibitem{Panahi2001}{ H. Panahi-Talemi and M. A. Jafarizadeh,}
{\em Procedings of the Third International Conference on Geometry,
Integrability and Quantization III,  Coral Press, Sofia,  269-381
(2001).}
\bibitem{Panahi2005}{ H. Panahi and M. A. Jafarizadeh,}
{\em J. Math. Phys. {\bf 46}, 012103 (2005).}
\bibitem{chaturvedi}{S. Chaturvedi and N. Mukunda,}
{\em Int. J. Mod. Phys. A {\bf 16}, 1481 (2001).}
\bibitem{akhtar1}{ S. J. Akhtarshenas,}
{\em J. Math. Phys. {\bf 48}, 012102 (2007).}
\bibitem{akhtar2}{ S. J. Akhtarshenas,}
{\em J. Phys. A: Math. Theor. {\bf 40}, 11333 (2007).}
\bibitem{georgi}{ H. Georgi,}
{\em ``Lie Algebras in Particle Physics''}, Addison-Wesley
Publishing Co., (1982).
\bibitem{gilmore}{ R. Gilmore,}
{\em ``Lie Groups, Lie Algebras, and Some of Their
Applications''}, John-Wiley Publishing Co., (1974).




\end{thebibliography}
\end{document}